\documentclass[journal]{IEEETran}
\usepackage{cite}
\usepackage{amsmath,amssymb,amsfonts}
\usepackage{algorithmic}
\usepackage{graphicx}
\usepackage[font=small,labelfont=bf,tableposition=top]{caption}
\usepackage{breqn}
\usepackage{subfigure}
\usepackage{multirow}
\usepackage{makecell}
\newcommand{\squeezeup}{\vspace{-5mm}}
\newcommand{\smallsqueezeup}{\vspace{-3mm}}
\usepackage[hidelinks]{hyperref}
\usepackage{graphicx}
\usepackage{xcolor}
\usepackage{textcomp}
\usepackage{stackengine} 
\DeclareUnicodeCharacter{2212}{-}
\newcommand\oast{\stackMath\mathbin{\stackinset{c}{0ex}{c}{0ex}{\ast}{\bigcirc}}}
\def\BibTeX{{\rm B\kern-.05em{\sc i\kern-.025em b}\kern-.08em
    T\kern-.1667em\lower.7ex\hbox{E}\kern-.125emX}}
\begin{document}
\setlength\intextsep{0pt}
\title{Simultaneous Denoising and Localization Network for Photoacoustic Target Localization }
\author{Amirsaeed Yazdani, Sumit Agrawal, Kerrick Johnstonbaugh, \emph{Student Member , IEEE}, Sri-Rajasekhar Kothapalli, \emph{Member, IEEE}, Vishal Monga, \emph{Senior Member, IEEE}
\thanks{We gratefully acknowledge the funding for this research
from NIH-NIBIB R00EB017729-05
(SRK), R21EB030370-01 (SRK), and Leighton Reiss Graduate Fellowship (SA)}
\thanks{Amirsaeed Yazdani and Vishal Monga are with the Electrical Engineering Department at Pennsylvania State University, State College, PA 16802, USA (email: amiryazdani@psu.edu, vum4@psu.edu)}
\thanks{Sumit Agrawal, Kerrick Johnstonbaugh, and Sri-Rajasekhar Kothapalli are with the Biomedical Engineering Department at Pennsylvania State University, State College, PA 16802, USA (email: sua347@psu.edu, kjohnstonbaugh97@gmail.com, szk416@psu.edu) }
% \thanks{Kerrick Johnstonbaugh is with the Biomedical Engineering Department at Pennsylvania State University,State College,PA 16802, USA (email:kfj5051@psu.edu) }
% \thanks{Sri-Rajasekhar Kothapalli is with the Biomedical Engineering Department at Pennsylvania State University,State College,PA 16802, USA (email:szk416@psu.edu) }
% \thanks{Vishal Monga is with the Electrical Engineering Department at Pennsylvania State University,State College,PA 16802, USA (email:vum4@psu.edu) }
}
\maketitle
\begin{abstract}
A significant research problem of recent interest is the localization of targets like vessels, surgical needles, and tumors in photoacoustic (PA) images. To achieve accurate localization, a high photoacoustic signal-to-noise ratio (SNR) is required. However, this is not guaranteed for deep targets, as optical scattering causes an exponential decay in optical fluence with respect to tissue depth. To address this, we develop a novel deep learning method designed to explicitly exhibit robustness to noise present in photoacoustic radio-frequency (RF) data. More precisely, we describe and evaluate a deep neural network architecture consisting of a {\em shared} encoder and two {\em parallel} decoders. One decoder extracts the target coordinates from the input RF data while the other boosts the SNR and estimates clean RF data. The joint optimization of the shared encoder and dual decoders lends significant noise robustness to the features extracted by the encoder, which in turn enables the network to contain detailed information about deep targets that may be obscured by noise. Additional custom layers and newly proposed regularizers in the training loss function (designed based on observed RF data signal and noise behavior) serve to increase the SNR in the cleaned RF output and improve model performance. To account for depth-dependent strong optical scattering, our network was trained with simulated photoacoustic datasets of targets embedded at different depths inside tissue media of different scattering levels. The network trained on this novel dataset accurately locates targets in experimental PA data that is clinically relevant with respect to the localization of vessels, needles, or brachytherapy seeds. We verify the merits of the proposed architecture by outperforming the state of the art on both simulated and experimental datasets. 
% These instructions give you guidelines for preparing papers for 
% IEEE Transactions and Journals. Use this document as a template if you are 
% using \LaTeX. Otherwise, use this document as an 
% instruction set. The electronic file of your paper will be formatted further 
% at IEEE. Paper titles should be written in uppercase and lowercase letters, 
% not all uppercase. Avoid writing long formulas with subscripts in the title; 
% short formulas that identify the elements are fine (e.g., "Nd--Fe--B"). Do 
% not write ``(Invited)'' in the title. Full names of authors are preferred in 
% the author field, but are not required. Put a space between authors' 
% initials. The abstract must be a concise yet comprehensive reflection of 
% what is in your article. In particular, the abstract must be self-contained, 
% without abbreviations, footnotes, or references. It should be a microcosm of 
% the full article. The abstract must be between 150--250 words. Be sure that 
% you adhere to these limits; otherwise, you will need to edit your abstract 
% accordingly. The abstract must be written as one paragraph, and should not 
% contain displayed mathematical equations or tabular material. The abstract 
% should include three or four different keywords or phrases, as this will 
% help readers to find it. It is important to avoid over-repetition of such 
% phrases as this can result in a page being rejected by search engines. 
% Ensure that your abstract reads well and is grammatically correct.
\end{abstract}

\begin{IEEEkeywords}
Deep Learning, Neural Networks, Photoacoustic Imaging, Optical Scattering, Target Localization.
% Enter key words or phrases in alphabetical 
% order, separated by commas. For a list of suggested keywords, send a blank 
% e-mail to keywords@ieee.org or visit \underline
% {http://www.ieee.org/organizations/pubs/ani\_prod/keywrd98.txt}
\end{IEEEkeywords}

\section{Introduction}
\begin{figure*}[h!]
    \centering
    \includegraphics[width=.75 \textwidth]{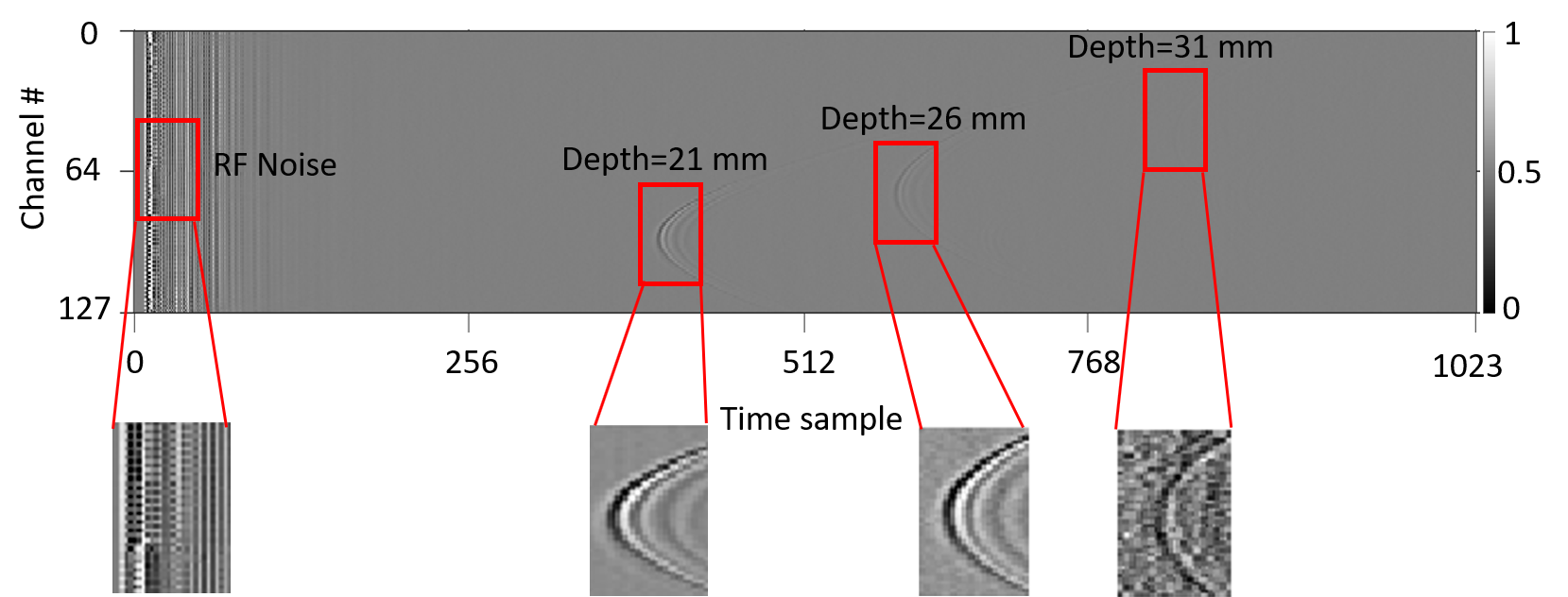}
    \caption{Experimental photoacoustic RF data shows the depth dependent signal strength from three $0.5\; mm$ diameter pencil lead targets, placed at $21\; mm$, $26\; mm$ and $31\; mm$ depths
    inside an optically scattering medium along with the RF noise.}
    \label{fig:fugure1}
\end{figure*}

Photoacoustic (PA) imaging is a promising method for the non-invasive visualization of optical contrasts in deep tissue. PA images (PAI) are typically reconstructed from measurements of acoustic waves generated by the thermoelastic expansion of light-absorbing molecules (e.g. hemoglobin) in the imaged tissue \cite{b38,b39}. Thermoelastic expansion of biological tissue occurs as a result of transient heating and expansion caused by the absorption of optical energy. Ultrasound transducers, capable of measuring high-frequency changes in pressure, capture the acoustic waves in the form of radio-frequency (RF) data. This data is typically reconstructed with a beamforming algorithm to produce a PA image representative of initial pressure distribution that is proportional to optical absorption in the tissue \cite{b44,b45,b46}. Given knowledge about the relative optical absorption of different tissue constituents, these images enable mapping of vascular structure and other tissue molecular contents. Moreover, the distribution of specific molecules and contrast agents (e.g. indocyanine green) can be determined, and blood oxygen saturation can be imaged \cite{b40,b41,b42,b43}. Despite these benefits, PAI suffers from low signal-to-noise ratios (SNR) when imaging deep tissue targets. This is mainly due to an unknown non-linear attenuation in optical fluence as a function of tissue depth. Figure \ref{fig:fugure1} provides a visualization of this effect in PA experimental RF data captured from three $0.5 \; mm$ diameter pencil lead targets, placed at different depths, in an optically scattering medium.

Deep learning has recently been shown to perform well in a  variety of problem domains encompassing image processing \cite{b1,b28,b29}, natural language processing\cite{b2,b3}, and other applications \cite{b62,b63,b64}. Specifically, Convolutional Neural Networks (\textbf{CNN}'s) have supplanted state of the art for image classification and segmentation with the ability to extract fine and coarse features from various types of images\cite{b4,b5,b6,b7,b8,b10,b36}. The most recent studies involving deep learning in the field of PAI benefit from CNNs as well. In the first significant effort towards PA target localization, Reiter \emph{et al}. in \cite{b9} map PA RF data to numerical coordinates using \textbf{VGG16} \cite{b6} followed by fully connected layers. Extending Reiter \emph{et al}.'s work, the authors in \cite{b11} use \textbf{faster R-CNN}\cite{b10} to propose regions for the targets as well as classify them as sources or artifacts. This work considers multiple point targets under different (simulated) noise levels; however, it doesn't consider the non-uniform and depth dependent optical fluence distribution inside tissue medium and associated with above-mentioned challenges. \color{black}The encoder-decoder structures like U-Net \cite{b4} are ideal for reconstruction tasks where the details in different fields of view are needed to be extracted by the encoder step by step through downsampling the input image. Moreover, the skip connections help the upsampling layers in the decoder generate the output with knowledge gained by layers in the encoder from lower fields of view. \color{black} Many past works benefited from this for sinogram super-resolution \cite{b57,b61}, and low CT denoising \cite{b50,b51,b52}. Of similar interest to denoising works, Lu et al. \cite{b59} propose a model based on generative adversarial network (GAN) for recovering high-quality optoacoustic images from limited-view images. Davoudi et al. \cite{b60} also make use of a U-Net based model to remove artifacts from optoacoustic  artifactual inputs. \color{black} To address the non-uniform optical fluence distribution, Johnstonbaugh \emph{et al}. \cite{b12,b13} proposed a structure based on U-Net with residual modules which downsamples the input data while extracting coarse detail. It subsequently upsamples the resulting low dimensional feature maps to construct a high-resolution heatmap and estimate PA target coordinates inside a deep tissue scattering medium. The network training  of Johnstonbaugh \emph{et al}. \cite{b12,b13} incorporates depth-dependent optical scattering but does not account for variable scattering levels of the tissue medium. Crucially, existing deep learning solutions for PAI target localization employ representative training RF image samples that are noisy but do not explicitly develop noise robustness by exploiting problem characteristics\footnote{There are other deep learning based works for denoising the PA images leading to improvement in imaging frame rates of PAI \cite{b47,b48}.}.
%This architecture performance degraded at higher optical scattering levels of 20 $cm^{-1}$ reduced scattering coefficient.  To address these issues we propose the following contributions: 

Recognizing that noise is a significant challenge in PAI, our work makes the following contributions:
\begin{itemize}
    \item \textbf{Novel Problem-Inspired Network Architecture}: We propose a \textbf{custom} architecture in which we extract noise-robust features from the input noisy images. Our proposal is a structure with a shared encoder and \textbf{two parallel decoders}. The first decoder is in charge of generating denoised images and therefore helping the encoder extract features that have vastly enhanced noise robustness. The second decoder localizes the targets in the input image using the features provided by the shared encoder. We, therefore, refer to our proposal as the \textit{Simultaneous Denoising and Localization Network (SDL).} \color{black} Our experimental results verify  that our proposed simultaneous strategy as opposed to a sequential or cascaded strategy (of denoising followed by localization) has state-of-the-art performance and  efficiency.  An  architecture  of  two-cascaded networks would be heavily parameterized, cumbersome to train and yields slower inference in the test phase.  For  the  same  number  of  network  parameters,  we  find (experimentally) that our SDL, in fact, leads to more accurate  localization. Moreover, our network also outperforms state of the art in terms of the denoising fidelity metrics.  \color{black}  
    
    \item \textbf{New  Regularized Training Loss Function}: While existing deep learning approaches in PAI \cite{b11,b12,b13} use a loss comprised of the differences between ground truth and network estimated target coordinates, we embellish our loss with new regularization terms that enhance noise robustness. First, because we perform denoising: a new fidelity prior/regularizer is used that optimizes the network parameters to minimize departures from the ground truth noiseless images used in training. Second, to incorporate the knowledge from the domain of photoacoustic imaging, we introduce processing on the estimated clean image via \textbf{wavefront/noise filters} to effectively help the network to focus on enhancing geometric attributes that help localize better while mitigating RF noise. These filters are designed based on the behavior of the noise and signal in the data and influence the network parameter optimization via new \textbf{domain-specific regularization} terms introduced in the training loss function.
    \item \textbf{A Practically Representative Simulated Dataset and Experimental Insights}: To comprehensively evaluate the effect of scattering level and the number of the targets, we generate a new dataset which is highly diverse with respect to the number of targets, their locations, and the background tissue scattering level. Our numerical evaluation shows greater localization accuracy and higher noise robustness than state of the art on both simulated and experimental data. We also investigate the effect of the number of training images (for the first time in the field of PA target localization) and show that as the number of training samples is reduced, the proposed SDL gracefully degrades in performance vs. competing methods, owing to the regularizers in SDL.
\end{itemize}

\begin{figure*}[h!]
    \centering
    \includegraphics[width=1 \textwidth, height=.6 \textwidth]{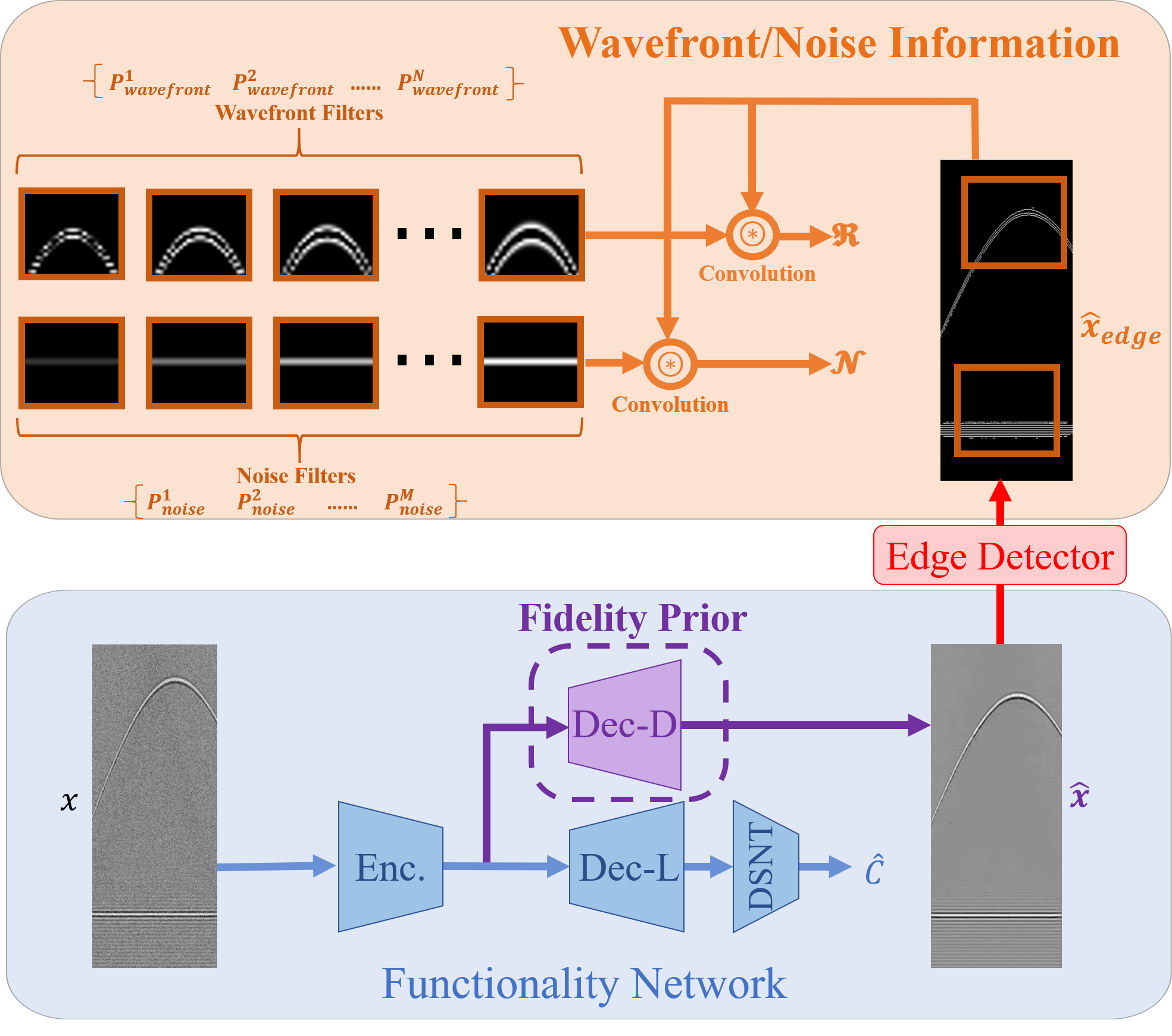}
    \caption{ The overall scheme of proposed SDL architecture. The \textbf{Functionality Network} consists of the main branches (encoder, denoising decoder, localization decoder, and differentiable spatial to numerical transform layer (DSNT)) to generate the coordinates and the denoised PA RF image. Note that the Dec-D branch essentially acts as a prior on the encoded features, which is enforced in the training via the regularizer in Eq. (\ref{eq:denoiser}). The \textbf{Wavefront/Noise Information} part calculates two terms needed for the regularizers in Eq. (\ref{wavefront}) and Eq. (\ref{noise}). This is accomplished by extracting edges from the denoised image and subsequently convolving them with noise and wavefront filters. The goal is to minimize the strength of the noise while matching the wavefront behavior to its counterpart in the ground truth noiseless image. Finally note that once trained, target localization involves processing the PA input image via the shared encoder and localization decoder (Dec-L).}
    \label{fig:network}
\end{figure*}
\section{Methods: Simultaneous Denoising and Localization Network}
\subsection{Joint Denoising-Localization Autoencoder}
\subsubsection{Encoder}
We use an autoencoder structure. Autoencoders have shown good performance, where features are extracted from images at various scales \cite{b13,b14,b15,b16,b17}. The encoder downsamples the input while constructing a low dimensional feature map, extracting low resolution, high field-of-view features. If we denote the input image by \textbf{x}, the encoder trainable parameters by $\theta_{enc}$, and the new representation of the input image in the lower dimensional feature space by \textbf{z}: $f(\mathbf{x},\theta_{enc})=\mathbf{z}$. The low dimensional feature map \textbf{z} is passed to upsampling modules in each of the two decoders.

\subsubsection{Localization Decoder}
The localization decoder  (Dec-L) upsamples the low dimensional input to the final size defined for the output heatmap. The heatmap represents the probability distribution of a target being located at any depth and lateral position. Finer features extracted in the initial layers of the encoder may be used (through skip connections) to refine the high-resolution heatmap. In terms of Dec-L parameters ($\theta_{Dec-L}$) and the feature map input (\textbf{z}), the heatmap (\textbf{h}) can be represented as follows: $ g_{1}(\mathbf{z},\theta_{Dec-L})=\mathbf{h}$. The numerical coordinates of PA targets are extracted using the  \textit{Differentiable Spatial to Numerical Transform (DSNT)} \cite{b18}.   
\subsubsection{Denoising Decoder}
So far, we have developed a typical autoencoder structure where no additional information from the noise present in the images is used. To do so, we use the features extracted by the encoder to estimate the clean RF input. In other words, we train a parallel decoder designed to output the high SNR version of the RF input: $g_{2}(\mathbf{z},\theta_{Dec-D})=\hat{x}$, where $\theta_{Dec-D}$ and $\hat{x}$ denote denoising decoder parameters and the cleaned RF data, respectively. The joint optimization of the shared encoder and dual decoders (denoising and localization) lends significant noise robustness to the features extracted by the encoder, enabling them to contain detailed information about deep targets obscured by noise.

 %output of this decoder is not directly used for localization. However, as we will define the loss terms in the following sections we notice that the encoder parameters will be affected through the optimization of this decoder, in other words joint optimization of the decoders indirectly couples them and leads us to more promising results.
% \begin{figure*}
%     \centering
    
%     \includegraphics[width=.8 \textwidth,height=.3 \textwidth]{Incomplete1.eps}
%     \caption{The scheme of proposed network.There are two paths from input to \textbf{numerical coordinates} and the \textbf{reconstructed clean image}.The latter is not involved in the test phase. }
%     \label{fig:general}
% \end{figure*}
\begin{figure*}
    \centering
    \includegraphics[width=1 \textwidth,height=.3 \textwidth]{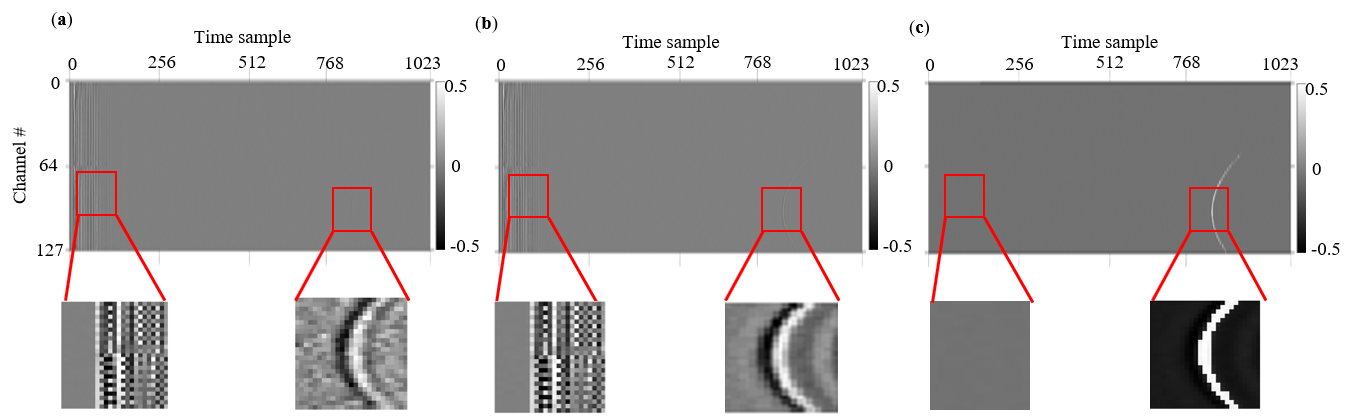}
    \caption{Depiction of how wavefront/noise filters are inspired. (\textbf{a}) Presence and intensity of signal wavefront and RF noise in PA noisy data. Expected clean outputs of the network (\textbf{b}) without wavefront and noise filter training, and (\textbf{c}) with wavefront and noise filter training. The goal is to suppress RF noise signals, while enhancing the strength of the PA signal wavefront.}
    \label{fig:new_fig}
\end{figure*}
\subsection{Feature Extraction and Optimization of the Network}
\subsubsection{Parallel Branches and Joint Optimization}As mentioned above, there are two paths in the network: \textbf{1)} Input $\rightarrow$ Heatmap and Numerical Coordinates and \textbf{2)} Input $\rightarrow$ Cleaned RF Data. The parameters in a neural network architecture must be optimized according to a loss function defined over the network outputs and desired ground truth via (typically) stochastic gradient descent\cite{b19} and backpropagation \cite{b20}. Our primary goal is to predict PA target coordinates as accurately as possible. Thus, the loss term defined on the output of the localization decoder is:
\begin{equation}
    \mathcal{L}(\theta_{enc},\theta_{Dec-L})=||\mathbf{C}-\hat{\mathbf{C}}||_{2}^{2}+\lambda_{1}\mathcal{JS}(\mathbf{h_{Gaussian}},\hat{\mathbf{h}})
    \label{eq:dec-L-loss}
\end{equation}
\color{black}where $\mathbf{C}$ and $\hat{\mathbf{C}}$ are both normalized Cartesian coordinates corresponding to the lateral and axial position of the target and the output of the network. (for example, $\mathbf{C}=(-1,-1)$  and $\mathbf{C}=(1,1)$ are the coordinates of the targets lying on the top left of the heatmap and bottom right of the heatmap, respectively.) \color{black}The operation $||.||_{2}^{2}$ denotes the $\ell _{2}$ norm. The second term imposes an additional constraint on the heatmap ($\hat{\textbf{h}}$), forcing it to follow a spherical Gaussian distribution ($\mathbf{h_{Gaussian}}$) by the measure of \textbf{Jenson-Shannon} divergence\cite{b18}.\\
We also intend to generate a high SNR version of the input RF data, evaluated with respect to ground truth clean images included in the training set. This is enforced by a fidelity prior/regularization term given by:
\begin{equation}
    \mathcal{L}_{Fidelity}(\theta_{enc},\theta_{Dec-D})=||\mathbf{x}_{clean}-\hat{\mathbf{x}}||_{2}^{2}
    \label{eq:denoiser}
\end{equation}
where $\mathbf{x}_{clean}$ is the ground truth high-SNR PA RF data. Eq. (\ref{eq:denoiser}) is used as a regularizer in the final training loss function
-- see Eq. (\ref{eq:finalLoss}). The joint optimization of the denoising decoder (Dec-D in Fig. \ref{fig:network}) with the shared encoder and localization decoder (Dec-L) in effect enhances target localization (i.e. the output of Dec-L). This is due to the fact that the encoder parameters (and hence resulting features) are optimized to simultaneously recover a clean PA image and accurately estimate target locations, which are tasks that benefit each other.\\
\textbf{Justification of the SDL Network Architecture in Fig.\ \ref{fig:network}}: The main benefit of a shared encoder-dual decoder architecture is that the denoising  decoder acts as a prior and ensures that the encoded features are noise-robust.  
The use of these noise-robust features then leads to superior results in target localization over networks that just use noisy images and ground truth coordinates for training.\\ 
An alternative option for using the knowledge of the noiseless images could be using a cascade of two networks in which the output of the denoising decoder (Dec-D) is first obtained and then fed into another deep network for
target localization.  We contend, however, that our proposed {\em simultaneous} strategy, as opposed to a {\em sequential or cascaded} strategy, has the upper hand both in  performance and efficiency. An architecture of two-cascaded networks would be heavily parameterized and cumbersome to train. For the same number of network parameters, we find (experimentally) that our SDL strategy, in fact, leads to more accurate localization. Second, the cascade of denoising and localization networks would be considerably slower than the proposed SDL for processing input PA image data. It is worth emphasizing that our proposed architectural enhancements in Fig.\ \ref{fig:network} only influence the training stage. Once trained, only the shared encoder and Dec-L are used to arrive at target coordinates, which means that SDL incurs {\em no additional} computational burden over existing state-of-the-art methods.
\begin{figure*}[h!]
    \centering

  \includegraphics[width=1 \textwidth, height=.4 \textwidth]{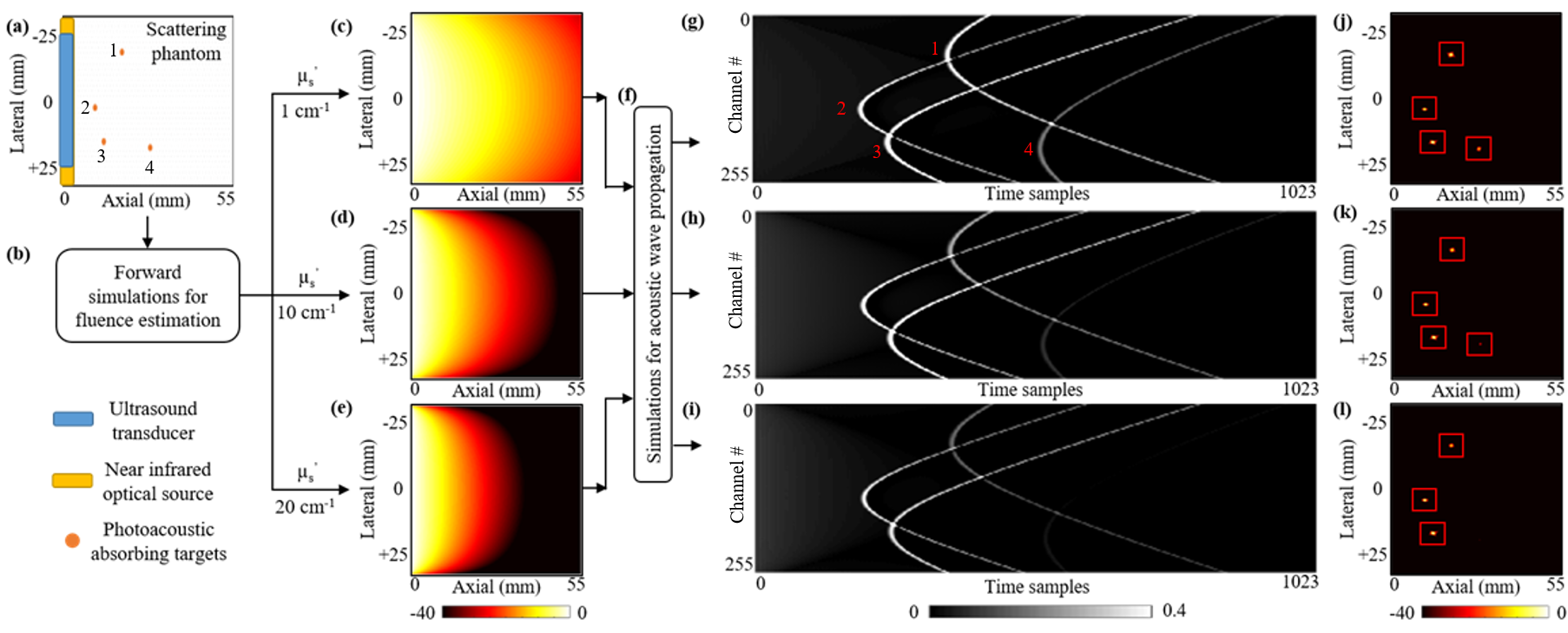}

    \caption{ The schematic of our simulated dataset generation platform with different optical scattering levels and numbers of targets. (\textbf{a}) A schematic view of the 55$\times$55 mm digital tissue phantom. A 51.2 mm wide 256-element ultrasound transducer array (blue stripe) and a 54 mm wide optical source (yellow stripe) are placed along the left edge of the phantom. Orange circles are  0.3 mm diameter vascular targets. (\textbf{b}) Simulated diffused light propagation via the \textbf{NIRFAST} toolbox solves optical fluence distributions for three different tissue mediums with reduced optical scattering coefficients of (\textbf{c}) 1 $cm^{-1}$; (\textbf{d}) 10 $cm^{-1}$; and (\textbf{e}) 20 $cm^{-1}$. (\textbf{f}) The photoacoustic wave propagation resulting from the optical fluence-induced initial pressure is simulated via the \textbf{K-Wave} toolbox. (\textbf{g}, \textbf{h}, \textbf{i}) The time sampled photoacoustic signals detected by the 256-element transducer array for each scattering level and (\textbf{j}, \textbf{k}, \textbf{l}) their corresponding beamformed images.}
    \label{fig:samples}
\end{figure*}
\subsubsection{Custom Regularizer and Wavefront/Noise Filters}
Fitting the denoising decoder output over the ground truth noiseless images can be achieved using the loss in Eq. (\ref{eq:denoiser}). However, this criterion may not lead to completely accurate outputs because, as we can see in Fig. \ref{fig:new_fig}, the RF data is mostly free of signal, and the $l_2$ norm in Eq. (\ref{eq:denoiser}) is a global measure that does not pose special emphasis on signal-containing regions in the data. This leads to gradients and updates that put significant weight on cleaning signal-free areas in the RF input. Moreover, even in the ground truth noiseless images, RF noise usually appears at the beginning or the end of data (Fig. \ref{fig:new_fig} a,b and Fig. \ref{fig:fugure1}), which we intend to mitigate. To achieve these objectives, we define new representation filters as depicted in Fig. \ref{fig:network}. These filters fall into one of two groups: \textbf{1) Wavefront filters}: which are generated using the idea of Scale and Curvature Invariant Ridge Detectors (SCIRD) \cite{b35} to extract the wavefronts (curves/signals) and \textbf{2) Noise filters}: which are crafted manually according to the behavior of RF noise in the images\footnote{RF noise may be removed using pre-processing steps on the PA image as well, but that process may not be perfect and in certain scenarios may, in fact, degrade or distort the signal too. Our proposed regularization will invariably help to enhance signal (wavefront) strength with respect to noise, albeit the extent of benefit may be dependent on the particular PA image data.}. The corresponding features are extracted by correlating these filters with the edges in the denoising network's cleaned RF outputs:
\begin{equation}
    \mathcal{R}^{i}=\{ P_{wavefront}^{i}\}_{i=1,..,N}\oast\hat{x}_{edge}
    \label{wavefront}
    \end{equation}
    \begin{equation}
    \mathcal{N}^{i}=\{P_{noise}^{i}\}_{i=1,...,M}\oast\hat{x}_{edge}
    \label{noise}
\end{equation}
Here, $P_{wavefront}$ and $P_{noise}$ denote wavefront and noise filters, respectively. $\hat{x}_{edge}$ represents the extracted edges in the cleaned output (extracted using simple edge filters) and $\oast$ is the convolution operation. The filters are defined based on the shape of the RF noise and signals in the images (see Fig. \ref{fig:network}). Thus, we have a new regularizer term defined as:
\begin{equation}
    \mathcal{L_{WN}}=\sum_{i=1}^{N}||\mathbf{\mathcal{R}_{clean}}-\mathbf{\mathcal{R}^{i}}||_{2}^{2}+\sum_{j=1}^{M}||\mathcal{N}^{j}||_{2}^{2}
    \label{WN}
\end{equation}
where $\mathbf{\mathcal{R}_{clean}}$ is the wavefront in the ground truth clean image. Thus, we impose a more restrictive constraint over the cleaned data by forcing it to follow the patterns in the ground truth, and at the same time remove the RF noise.\\
The loss function to be minimized for the whole network is the sum of the central loss term in Eq. (\ref{eq:dec-L-loss}), the fidelity prior/regularizer in Eq. (\ref{eq:denoiser}), and regularization term involving wavefront and noise filters in Eq. (\ref{WN}):
\begin{multline}
\label{eq:finalLoss}
    \mathcal{L}(\theta_{enc},\theta_{Dec-L},\theta_{Dec-D})=||\mathbf{C}-\hat{\mathbf{C}}||_{2}^{2}+\\\lambda_{1}\mathcal{JS}(\mathbf{h_{Gaussian}},\hat{\mathbf{h}})+\lambda_{2} \mathcal{L}_{Fidelity} + \gamma\mathcal{L_{WN}}
\end{multline}
where $\lambda_{1}$, $\lambda_{2}$ and $\gamma$ are regularizer constants for Jenson-Shannon divergence, the fidelity, and wavefront/noise loss, respectively. Note that the regularization constants are experimentally determined by cross-validation \cite{ b49}. $N$ and $M$ denote the number of wavefront and noise filters, respectively.

\subsection{New Practically Representative Simulated Dataset}
% \label{sec:guidelines}

For the success of any deep learning application in the medical imaging domain, the problem of obtaining large training datasets must be addressed. Conventionally, the researchers in PAI community have used various types of simulations to obtain the training data. Allman \emph{et al}. \cite{b11} have used the K-wave simulation tool to obtain the photoacoustic raw data, considering a uniform optical fluence along the tissue depth. These simulations are helpful while considering the illumination source always present near the target, which is true for endoscopic applications. However, for deep tissue non-invasive imaging conditions, heterogeneous optical fluence distribution must be considered. To address the problem of PA signal attenuation due to scattering, Agrawal \emph{et al}. \cite{b30} recently introduced a hybrid simulation platform where depth-dependent fluence attenuation is considered while obtaining the photoacoustic raw data. Johnstanbaugh \emph{et al}. \cite{b12} used this simulation platform to obtain a large training dataset of single target photoacoustic images with the scattering of background tissue  fixed to 10 $cm^{-1}$, in contrast to realistic PAI scenarios where the scattering noise levels might vary up to 20 $cm^{-1}$. In the work presented here, we have addressed these shortcomings by using the same simulation platform \cite{b30} to produce a dataset that includes multiple point targets placed at different depths inside a deep tissue medium with three different $\mu_{s}' = $1 $cm^{-1}$, 10 $cm^{-1}$, and 20 $cm^{-1}$. As shown in Fig. \ref{fig:samples}, this hybrid simulation platform uses the NIRFAST toolbox to solve for the optical fluence distribution inside the tissue medium
\cite{b31,b32}. At each point in the digital phantom, the calculated optical fluence is multiplied by the optical absorption coefficient at that point to yield the initial pressure distribution. The K-Wave function \textit{kspaceFirstOrder2D} \cite{b33} takes the initial pressure distribution as input and simulates the propagation and detection of the resulting photoacoustic waves. The platform outputs time vs. pressure measurements detected with a 256-element ultrasound transducer array.\\
Using this platform, simulated PA signal measurements were generated for 8000 digital (in silico) tissue phantoms. The digital phantom consists of a 276×276 two-dimensional grid with 0.2 mm node spacing (Fig. \ref{fig:samples}a). This grid represents a 55 mm$\times$55 mm sized soft tissue. For each phantom, an integer from 1 to 4 was randomly selected. This number of blood targets was placed at random positions within the 10 mm to 50 mm depth range of the homogeneous tissue medium. The lateral positions of the vascular targets were constrained to a range of 10 mm on either side of the center of the transducer array. After the targets were placed, PA signal propagation was simulated with three different background scattering levels ($\mu_{s}'=$1 $cm^{-1}$, 10 $cm^{-1}$, and 20 $cm^{-1}$) (Figs. \ref{fig:samples}g, h, i).
\subsubsection{Optical Fluence Calculation} The first step in generating a photoacoustic signal is the photoexcitation of optically absorbing molecules. Acoustic wave propagation follows thermoelastic expansion of such molecules, and the initial pressure is modeled as proportional to the local optical fluence (Eq. (\ref{eq:proportional})). We adopt the diffusion approximation model of light propagation \cite{b31} in our digital scattering phantoms (Fig. \ref{fig:samples}b):
\vspace{-1.5ex}
\begin{equation}
    -\nabla \cdot \kappa(r)\nabla\phi(r,\omega)+\Big(\mu_a(r)+\frac{i\omega}{c_m(r)}\Big)\phi(r,\omega)=q_0(r,\omega)
\end{equation}
where $\phi(r, \omega)$ is the fluence rate at position r and modulation frequency $\omega$, $\mu_{a}$ and $\mu_{s}$ are the absorption and reduced scattering coefficients, the diffusion coefficient $\kappa = 1/3(\mu_{a} +\mu_{s})$, $q_{0}(r, \omega)$ is an isotropic optical source term, and $c_{m}(r)$ is the speed of light at position r in the tissue.
 \begin{figure}
        \centering
        \includegraphics[width=.5\textwidth]{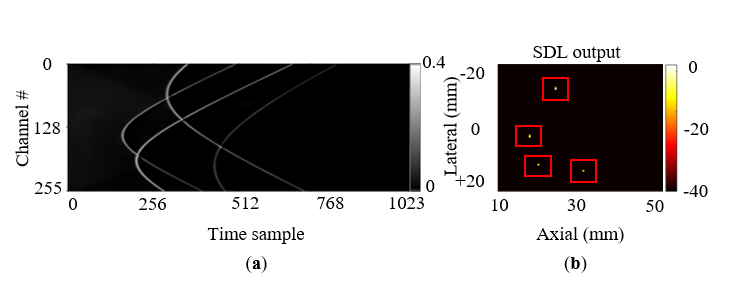}
        % \subfigure[]{
        % \includegraphics[width=.5\textwidth ,height=.15 \textwidth]{input_sim2_perfect.eps}}\\
        % \subfigure[]{
        % \includegraphics[width=.5\textwidth, height=.15 \textwidth]{cleaned_sim2_perfect.eps}}\\
        % \subfigure[]{
        % \includegraphics[width=.15\textwidth, height=.15 \textwidth]{heat_sim2_perfect.eps}}
        % \includegraphics[width=.3 \textwidth ,height=.3 \textwidth]{heat_exp2_perfect.eps}\\
        % \includegraphics[height=.15 \textwidth ,width=.15 \textwidth]{combined_sim2_perfect.png}
        \caption{(\textbf{a}) Denoised output and (\textbf{b}) the corresponding heatmap of 4 photoacoustic targets used in Fig. \ref{fig:samples}i (as it will be explained in section III.D) Note that SDL successfully detects all 4 targets, whereas one is missed by conventional beamforming approaches in Fig. \ref{fig:samples}i.
        }
        \label{fig:sim2}
    \end{figure}
    
\subsubsection{Initial Pressure Calculation and Propagation}The initial pressure distribution in the phantom is calculated by multiplication of the optical fluence at each point with the corresponding optical absorption (Fig. \ref{fig:samples}f):
\begin{equation}
    p_{0}(r)=\Gamma\mu_{a}(r,\lambda)\phi(r)
    \label{eq:proportional}
\end{equation}
where $\phi(r)$ and $\mu_{a}(r, \lambda)$ are the optical influence at position r and optical absorption coefficient at position r and excitation wavelength $\lambda$, respectively. $\Gamma$ is the Gr\"uneisen parameter, which we model as equal to one unit for simplicity. $p_{0}$ is the initial pressure at tissue position r. The value of the optical absorption coefficient of oxygenated human blood is modeled with an optical absorption coefficient of $\mu_{a} = 0.425$ $mm^{−1}$ at 800 nm \cite{b34}. The initial pressure distribution is input to the k-Wave function \textit{kspaceFirstOrder2D}, which simulates the propagation of acoustic waves in the tissue medium (Fig. \ref{fig:samples}d). The acoustic propagation is solved for using the following set of equations \cite{b34}:
\begin{equation} \label{eqn:acoustic_prop}
    \frac{\partial \textbf{u}}{\partial t}=-\frac{1}{\rho_0}\nabla p, \;
    \frac{\partial \rho}{\partial t}=-\rho_0 \nabla \cdot \textbf{u}, \;
    p=c^2\rho
\end{equation}
Where $p$ is the acoustic pressure, \textbf{u} is the particle velocity, $\rho_{0}$ is the ambient density in the tissue, $\rho$ is the acoustic density, $c$ is the tissue speed of sound.
\begin{figure*}
    \centering
    \includegraphics[width=1 \textwidth]{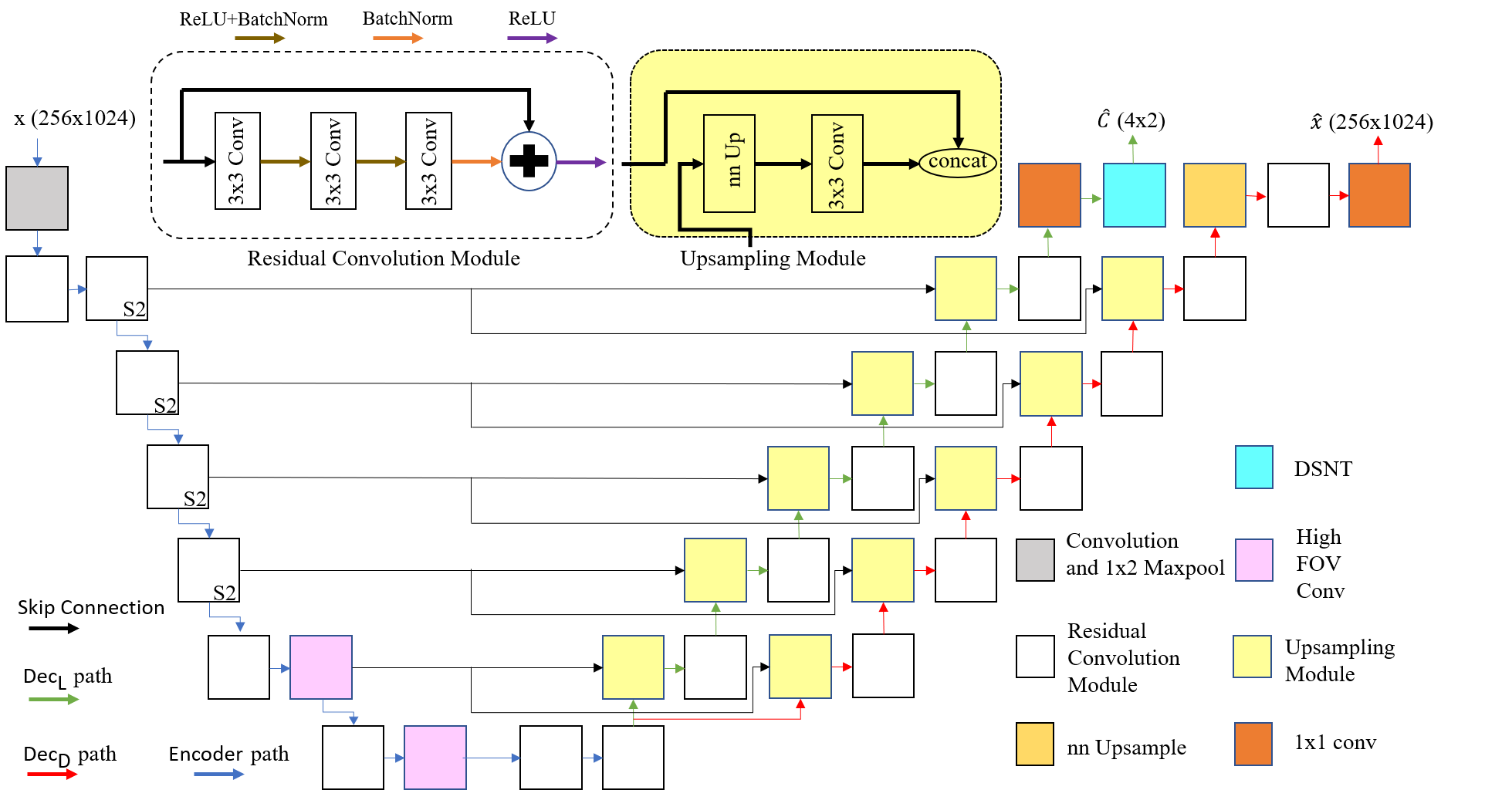}
    \caption{\color{black}The architecture of the U-Net based network  used for the encoder and decoders. Different colored boxes represent the following specific modules: Convolution and 1x2 Maxpool, residual convolution module (S2 on the right corner of some modules indicates that they downsample the input features with stride 2), nn Upsample which is based on the nearest neighboring algorithm, DSNT layer, Upsampling module, 1x1 conv, and high field-of-view convolution module with 5x5 kernels and strides 1 and 3. They are termed the high field-of-view modules because the size of kernels (5x5) is comparable to the size of feature maps (16x32) which is useful for detecting the dependencies between far apart pixels in the input image. }
    \label{fig:net_arch}
\end{figure*}
\section{Experimental Results}\subsection{Network Implementation} 

\noindent \textbf{Reproducibility}: Our implementation code for different training/test scenarios, data files for wavefront/noise filters, training, and test dataset, and the exact package numbers can be found at: \href{https://github.com/yazdaniamir38/SDL-network-for-PA-target-localization}{\underline https://github.com/yazdaniamir38/SDL-network-for-PA-target-localization}. \\
Our network is implemented using \textbf{Python} and deep learning packages of \textbf{Pytorch} \cite{b21} and all of the training and testing experiments were conducted on an NVIDIA Titan X GPU (12GB). \color{black} We trained our model for 100 epochs with batch size 4 using \textbf{RMSprop} \cite{b53} optimizer with an initial learning rate of 0.0001. We used a learning rate scheduler with a decrease factor of 0.1 and a patience factor of 10 epochs. To find the optimum values for hyperparameters including regularizing constants, we performed cross-validation by keeping a portion of training data as the validation set (2000 samples) and training the network for a couple of epochs to see how fast and stably the error decreases. The values we found by cross-validation for $\lambda_{1}$, $\lambda_{2}$, and $\gamma$ are 1, 0.5, and $10^{-4}$ respectively. The simultaneous presence of multiple regularization terms may prevent the training loss of the network from converging. So we initially train the model without the $\mathcal{L_{WN}}$ term. After about 60 epochs when the model converges to a suboptimal point, we activate the terms corresponding to $\mathcal{L_{WN}}$ to further refine the network parameters. We extract windows of size 45x40 pixels centered on the wavefront in noiseless samples as $\mathcal{R}_{clean}$ in Eq. (\ref{WN}). To approximately estimate the location of the wavefront in the image, we make use of the ground truth coordinates and project them to pixel coordinates. The encoder and localization decoder architecture (see Fig. \ref{fig:net_arch} for details) follow \color{black}the residual U-Net based structure of \cite{b12,b13}  with changes in the number of the output channels of the last convolutional layer according to the number of targets to be detected. As we will further explain in different setups, the network can be used for localizing either single targets or a variable number of targets. The denoising decoder jointly with the shared encoder also follow the residual U-Net architecture (with a different number of layers) \color{black}and consist of the following:
\subsubsection{Residual Convolutional Modules} These modules consist of three convolutional blocks with a 3$\times$3 kernels where each is followed by a batch normalization layer and a ReLU activation function. The output of the last batch normalization layer of each module is added to the input of the same module and passed to a ReLU activation function. The stride in the first block is 2 to downsample the input. The purpose of these modules is to encode features at various scales. Downsampling enables larger-scale features to be captured with 3x3 convolutions via hierarchical feature extraction. \color{black} Fig. \ref{fig:net_arch} depicts these modules with white boxes and S2 on the right corner of the boxed denotes that the input features are downsampled with stride 2.\color{black}
\subsubsection{Upsampling Modules} These modules increase the resolution of  the low dimensional features  to the desired size of the output. There is a nearest neighbor upsampling block, with a factor of 2, to upsample the input followed by a convolutional block with 3$\times$3 kernel. The output of the convolutional block is concatenated with the corresponding skip connections from the encoder and passed to a residual module. \color{black}In Fig. \ref{fig:net_arch}, these modules are shown in yellow and fed with two inputs.\\\color{black}
Since the output of the denoising decoder (cleaned output) is larger than the output of the localization decoder (heatmap), the denoising decoder has one extra module. \cite{b13} uses the \textbf{Nyquist} convolutional layer to downsample the image; however, in the presence of the noise,  better performance is observed, when the input image is not downsampled. In total, the encoder, the localization decoder, and the denoising decoder have 7, 6, and 7 layers, respectively. Table \ref{tab:network} shows the details of each layer in the encoder and the decoders \color{black} with respect to input/output size and Fig. \ref{fig:net_arch} shows the model architecture. \color{black}
% The encoder is mainly composed of residual convolution modules and 2 high Field of view convolution modules. The residual modules provide the next layer the extracted features along with features they get from the previous layer, this would help for extracting more generalised features. Moreover, the skip connections provide the additional information the upsampling modules in the decoder need for upsampling the downsampled input to the objective size of either the denoised output or the the heatmap.   \color{black} 
\begin{table*}
  \caption{The details for the layers of the \textbf{encoder} ($Enc$), \textbf{Localization decoder} ($Dec-L$), and \textbf{Denoising decoder} ($Dec-D$). The size of the PA input is 256*1024. The definition of encoder in our work is slightly different with \cite{b13}'s as we consider layer 7 belonging to the encoder and its output is considered as the shared input for decoders. Each layer for decoders consists of a residual and upsampling module except for the last layer which is a simple convolutional layer and all of these layers benefit from the skip connections of their corresponding layer in the encoder.}
    % \centering
    \resizebox{\textwidth}{!}{
  
    \begin{tabular}{|c|c|c|c|c|c|c|c|c|}
    \hline
    Layers&Number of Kernels/ &Size of output&Layers&Number of Kernels/&Size of output &Layers&Number of Kernels/&Size of output\\ 
    &feature maps&feature maps&&feature maps&feature maps&&feature maps&feature maps\\
    \hline
         $Layer1_{Enc}$&16&256*1024&$Layer8_{Dec-L}$&256&16*32&$Layer8_{Dec{D}}$&256&16*32\\
         $Layer2_{Enc}$&16&256*512&$Layer9_{Dec-L}$&128&32*64&$Layer9_{Dec-D}$&128&32*64\\
         $Layer3_{Enc}$&32&128*256&$Layer10_{Dec-L}$&64&64*128&$Layer10_{Dec-D}$&64&64*128\\
         $Layer4_{Enc}$&64&64*128&$Layer11_{Dec-L}$&32&128*256&$Layer11_{Dec-D}$&32&128*256\\
         $Layer5_{Enc}$&128&32*64&$Layer12_{Dec-L}$&16&256*512&$Layer12_{Dec-D}$&16&256*512\\
         $Layer6_{Enc}$&256&16*32&$Layer13_{Dec-L}$&1&256*512&$Layer13_{Dec-D}$&8&512*1024\\
         $Layer7_{Enc}$&256&8*16&&&&$Layer14_{Dec-D}$&1&256*1024\\
         \hline
    \end{tabular}}
    \label{tab:network}
\end{table*}
\vspace{-3mm}
\subsection{Experimental Setup}
The performance of the proposed network is compared with the state of the art by considering different experimental setups. Furthermore, we designed new scenarios to evaluate the network in more challenging conditions and find the boundary to which our network can still show good performance. Our different scenarios include \textbf{i) comparing the performance of our network over the datasets employed in \cite{b11}} and \textbf{ii)\cite{b13} for a similar application, iii) performance on our own practically representative simulated  dataset and experimental phantoms with multiple targets, including inside chicken breast tissue.} Unless otherwise stated, for the proposed SDL -- the number of 
wavefront filters $N = 10$ and the number of noise filters was set to $M = 20$. These numbers were determined by cross-validation for the best overall results.
\smallsqueezeup
\subsection{Comparisons with State of the Art}
\subsubsection{Experiments on Allman \emph{et al}.'s Dataset: Up to Two Targets} 
\begin{figure*}
       \centering
       \includegraphics[width=1 \textwidth]{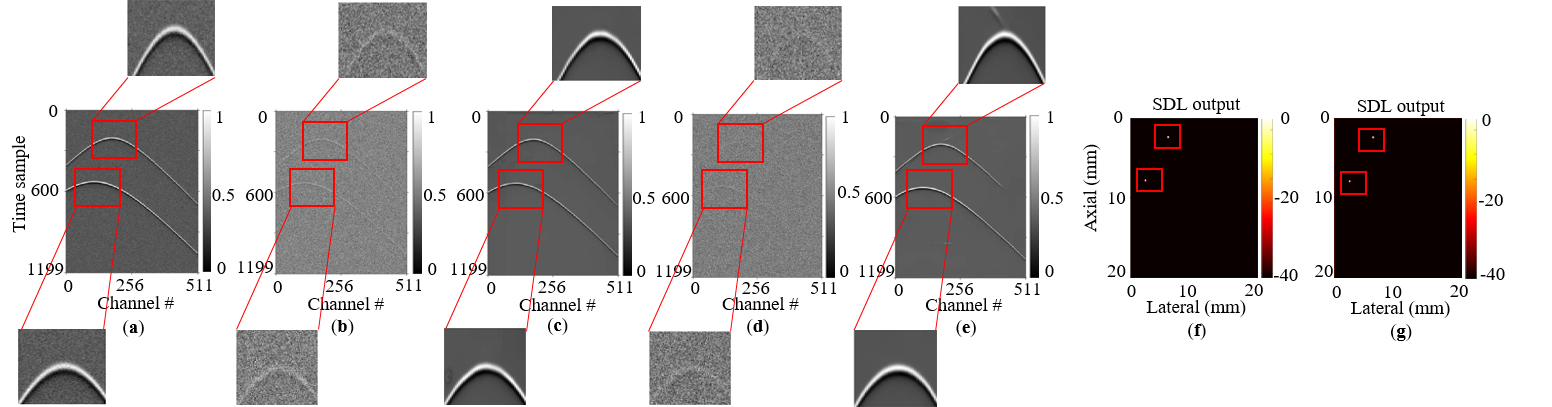}
    %   \subfigure[]{
    %   \includegraphics[height=.2 \textwidth,width=.12 \textwidth]{TMI_noiseless_perfect.eps}}
    %     \subfigure[]{
    %   \includegraphics[height=.2 \textwidth,width=.12 \textwidth]{cleaned_-3_perfect.eps}}
    %     \subfigure[]{
    %   \includegraphics[height=.2 \textwidth,width=.12 \textwidth]{cleaned_-9_perfect.eps}}
    %     \subfigure[]{
    %   \includegraphics[height=.2 \textwidth,width=.12 \textwidth]{TMI_-3_perfect.eps}}
    %     \subfigure[]{
    %   \includegraphics[height=.2 \textwidth,width=.12 \textwidth]{TMI_-9_perfect.eps}}
    %     \subfigure[]{
    %   \includegraphics[height=.2 \textwidth,width=.12 \textwidth]{heat_-3_perfect.eps}}
    %     \subfigure[]{
    %   \includegraphics[height=.2 \textwidth,width=.12 \textwidth]{heat_-9_perfect.eps}}
       \caption{Performance of our SDL network on \cite{b11}'s photoacoustic dataset. (\textbf{a}) A noiseless sample. (\textbf{b},\textbf{c}) Its -3 dB noisy version and its reconstructed output from our network, respectively. (\textbf{d},\textbf{e}) -9 dB noisy version and its reconstructed output, respectively. (\textbf{f},\textbf{g}) Output heatmaps generated by the network for -3 dB and -9 dB inputs, respectively. }
       \label{fig:TMI}
        \squeezeup
   \end{figure*}

\color{black}To perform a fair comparison with the work of  \cite{b11}, we use their released simulated dataset\footnote{https://ieee-dataport.org/open-access/photoacoustic-source-detection-and-reflection-artifact-deep-learning-dataset}. This dataset contains 16,000 and 4,000 images for training and testing, respectively, with at most two targets (source and artifact). The labels provided for the targets are the bounding box coordinates, whose centers are used as labels for SDL. \color{black}We compare the networks in 3 different scenarios: 1) noiseless images, 2) noisy images when the signal to noise ratio (\textbf{SNR}) is -3dB, and 3) when the SNR is -9dB. Zero mean Gaussian noise with a certain standard deviation is added to obtain the desired noise levels and the noiseless images are used as ground truth for our denoising decoder. Of note, since in this dataset scattering noise level is not considered, the signal intensity in different depths is almost fixed. Table \ref{tab:Comparison} compares the results for both networks. Table \ref{tab:Comparison} confirms that SDL reduces the error significantly over \cite{b11}, particularly for the -3 dB case. Some of the test samples along with their corresponding cleaned outputs and heatmaps from our network are shown in Fig. \ref{fig:TMI}. It may be observed that SDL predictions have higher precision, since the network generates a very small concentrated point versus a bounding box.  
 \begin{table}
%  \scriptsize
 \caption{Comparing the results in the same manner, \cite{b11} reported the results over their simulated test set. \color{black}T is the inference time reported per image sample.}
  \label{tab:Comparison}
  \centering
  \
    \begin{tabular}{|c|c|c|c|c|}
      \hline 
     \multirow{2}{*}{Architecture}   &\multicolumn{3}{c|}{percentage of total Error$\leq$0.5 mm}&\multirow{2}{*}{\color{black} T(sec)} \\ 
      \cline{2-4}
          &Noiseless&  SNR$=-3dB$& $SNR=-9dB$&\\
      \hline
    \cite{b11} &93.73&96.51&95.63& \color{black} 0.068 \\
    \hline
     SDL& \textbf{99.71}&\textbf{99.47}&\textbf{96.65}&\color{black}\textbf{0.058} \\
      
  \hline
    
  \end{tabular}
  \squeezeup
   \end{table}
   
\subsubsection{Experiments on Johnstonbaugh \emph{et al}.'s Dataset: Single Target with Tissue Optical Scattering}
The dataset used by Johnstonbaugh \emph{et al}.\cite{b13} is much more realistic with regard to scattering noise level, but  only considers the single target localization within a deep tissue medium of fixed background scattering value. \color{black}This dataset contains 16,240 and 4,060 samples for training and test, respectively. Since we are using their proposed autoencoder structure as a base for our encoder and denoising decoder, we have the same performance in the case without additive Gaussian noise. To obtain the objective SNR (-9 dB), the average signal intensity in depth 40 mm is calculated and the variance of the noise is defined according to that. \color{black}The numerical results for both networks are compared in Table \ref{Comparison_tuffc} \color{black}where lateral, axial, and Euclidean error are calculated as follows:
\begin{multline}
 Error_{Ax}=|x-\hat{x}| \hspace{7ex}
	    Error_{Lat}=|y-\hat{y}|  \\ 
	    Error_{Euc}=\sqrt{Error_{Lat}^{2}+Error_{Ax}^{2}}
\end{multline}
where $x$ and $\hat{x}$ denote the ground truth and network's output for axial coordinate, respectively.  $y$ and $\hat{y}$ are the ground truth and network's output for lateral coordinate, respectively. \color{black}These results indicate that injecting PA domain knowledge about RF and scattering dependent noise - as SDL exploits -- improves the results specifically for the targets that lie deeper (than 40 mm) in the scattering noise.
\begin{table}
\caption{Performance over \cite{b13}'s dataset with respect to depth. We can see the significant improvement over deep targets which is due to the effect of denoising part as deeper targets suffer from  optical scattering more.}
\centering
  \label{Comparison_tuffc}
  \resizebox{.5 \textwidth}{!}{
    \begin{tabular}{|c|c|c|c|}
      \hline 
    \multirow{ 2}{*}{Architecture}     & \multicolumn{3}{|c|}{10 mm$\leq$ Depth$\leq$50 mm}\\
    % \hline
    &Axial ($\mu m$)&Lateral ($\mu m$) & Euclidean($\mu m$)\\
      \hline
   \cite{b13}& 26.95&116.43& 125.51 \\
     SDL &\textbf{7.62}&\textbf{77.64} &\textbf{81.21}  \\
     \hline
    \multirow{ 2}{*}{Architecture}     & \multicolumn{3}{|c|}{Depth$>$40 mm}\\
   & Axial ($\mu m$)&Lateral ($\mu m$) & Euclidean($\mu m$)\\
      \hline
    \cite{b13}&82.24&427.11&450.99\\
     SDL &\textbf{21.07} & \textbf{252.65}&\textbf{256.42}\\
  \hline
\end{tabular}}
\squeezeup
  \end{table}
  
\begin{figure}
    \centering
     \includegraphics[width=.5 \textwidth ]{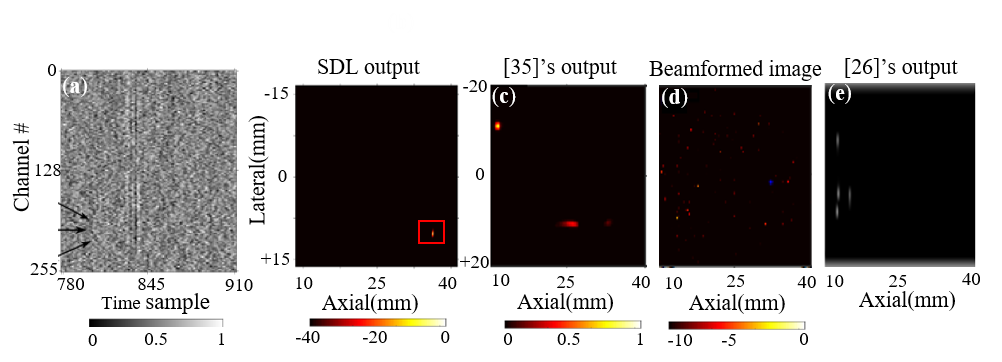}
    \caption{\color{black}Comparison between SDL,\cite{b11}, \cite{b13}'s performance, and traditional beamforming. (\textbf{a}) Experimental RF data with 20 $cm^{-1}$ scattering noise taken from \cite{b13}, (\textbf{b}) the prediction from SDL, (\textbf{c}) \cite{b13}'s output, (\textbf{d}) beamformed image, and (\textbf{e}) \cite{b11}'s output (bounding box centers are shown).}
    \label{fig:exp1}
\end{figure}

\noindent Johnstonbaugh et al.\cite{b13} also test the network on experimental samples with different scattering levels and in the case of 20 $cm^{-1}$ noise level, as shown in Fig. \ref{fig:exp1}, their network fails (\color{black}as well as the network proposed in \cite{b11}). \color{black}The output of our network (after it's trained with their modified dataset and tested on these samples) can be seen in Fig. \ref{fig:exp1}, where SDL can still predict even with the extreme level of scattering.
\begin{figure}[ht]
\setlength\belowcaptionskip{-0.7\baselineskip}
    \centering
    \includegraphics[width=.4 \textwidth]{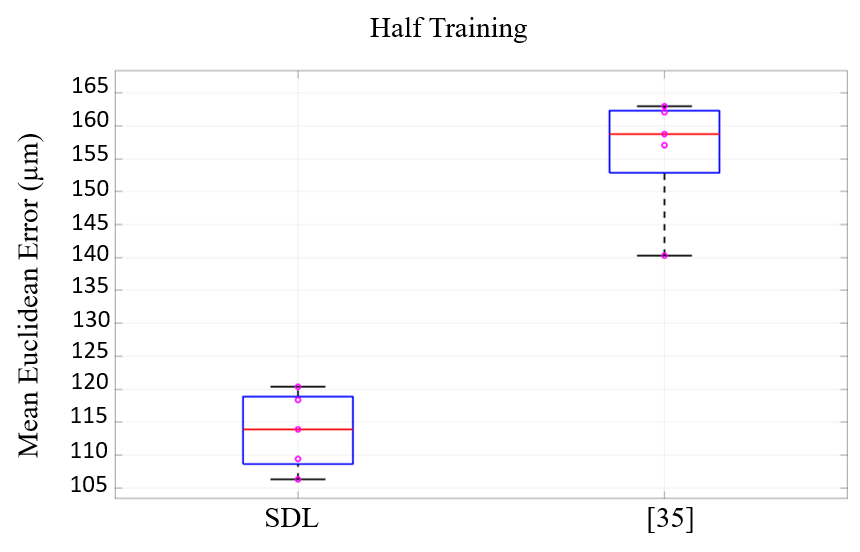}
    \caption{\color{black}The performance results of the models for 5 different training/test splits when the training size reduces to half (8,120 samples). Our lower mean and variance shows the reliability of the network in limited training regime.}
    \label{fig:chart}
    \smallsqueezeup
\end{figure}

\noindent\textbf{Effect of training sample size:} \color{black}
To observe the effect of priors and carefully devised architectures, it's common in state-of-the-art methods \cite{b54, b55,b56} to evaluate the performance of the models when the training set is limited. This is important from a practical point of view as in most of real world cases the number of annotated samples is limited. Therefore networks should be able to maintain their performance with the least degradation possible compared to when training samples are not limited. The models should also show robustness when the train/test splits change. To investigate this, we chose 5 different randomly chosen train/test splits with a training size of 8,120 and a test size of 4,060 using the original dataset with SNR=-9dB (used in \cite{b13}). We trained SDL and the model in \cite{b13} 5 times on different training sets and tested them on the corresponding test set. Fig. \ref{fig:chart} conveys two key messages: 1) the average error for SDL even when training is halved is still quite good and much better than what \cite{b13} achieves with half the data; 2) the significantly lower variance of the error for SDL vs. \cite{b13} for the 5 train/test splits shows that SDL is much more robust to \textit{selection bias}, i.e. SDL performance is less sensitive to the exact choice of training samples - an important practical benefit, which is provided by the priors/regularizers used with SDL.\color{black}
\smallsqueezeup
\subsection{Experiments on Challenging Simulated and Real Data}
We use the dataset in Section II-C for evaluating the capability of the network with respect to the noise level. To do so, we trained the network 2 times, over 2 different training sets of size 8,000, where the scattering noise level in each set was different (10 $cm^{-1}$ and 20 $cm^{-1}$). The samples with 1 $cm^{-1}$ noise level were used as ground truth for the denoising decoder. Each trained model was tested on a test set with the size of 2000 and corresponding scattering level. 
\color{black}In addition to separate noise level experiments, we also trained the SDL network over a large training set obtained by varying the scattering noise level (1 $cm^{-1}$, 10 $cm^{-1}$, and 20 $cm^{-1}$) and tested it over the combined test set -- these are labeled as {\em mixed} in Table \ref{tab:new_dataset}. Table \ref{tab:new_dataset} shows the results with respect to depth and distinct noise levels. Results in Table \ref{tab:new_dataset} reveal that as expected, the average errors for SDL are generally higher for the higher scattering noise level. We should mention that average localization errors for the {\em mixed} case in Table \ref{tab:new_dataset} appear lower because the training set is a combined set of all the training samples with different scattering coefficients and hence it's richer; moreover, the test includes the 1 $cm^{-1}$ case, which is practically noiseless.

\begin{figure*}
    \centering
    \includegraphics[width=.9 \textwidth]{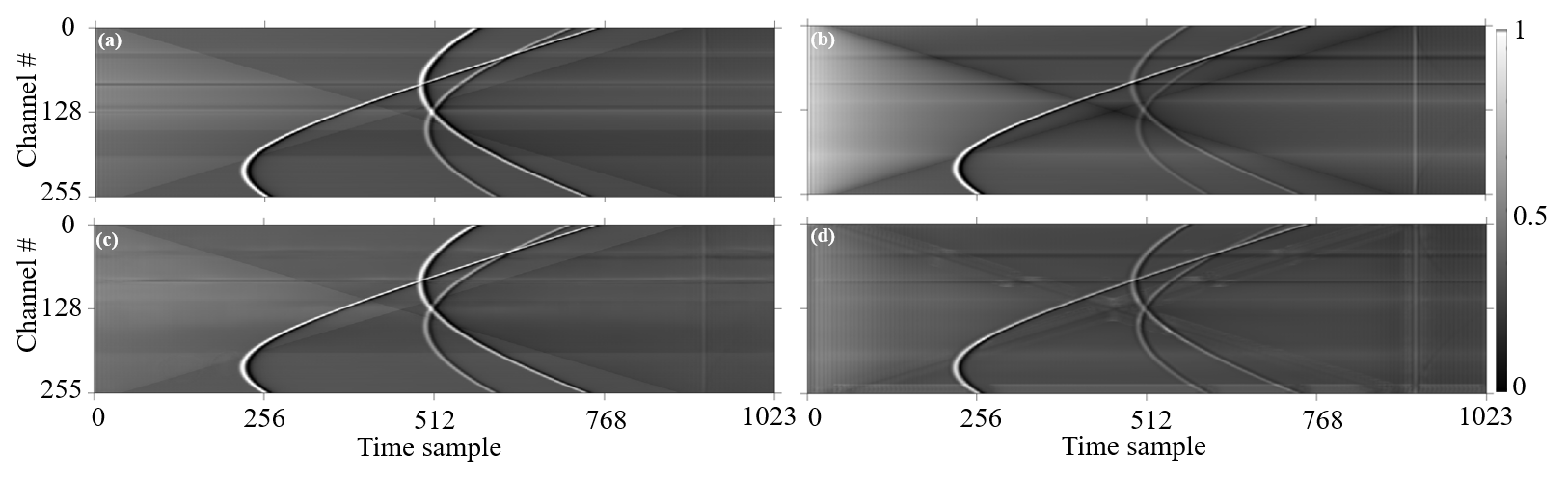}
    \caption{\color{black}The denoising performance of SDL and the cascaded network on a test sample: (\textbf{a}) noiseless ground truth, (\textbf{b}) noisy input, (\textbf{c}) SDL denoised output, and (\textbf{d}) the cascaded network denoised output.}
    \label{fig:denoise}
\end{figure*}

\begin{figure*}[h!]
    \centering
% \subfigure[]{
    \includegraphics[width=1 \textwidth,height=.24 \textwidth]{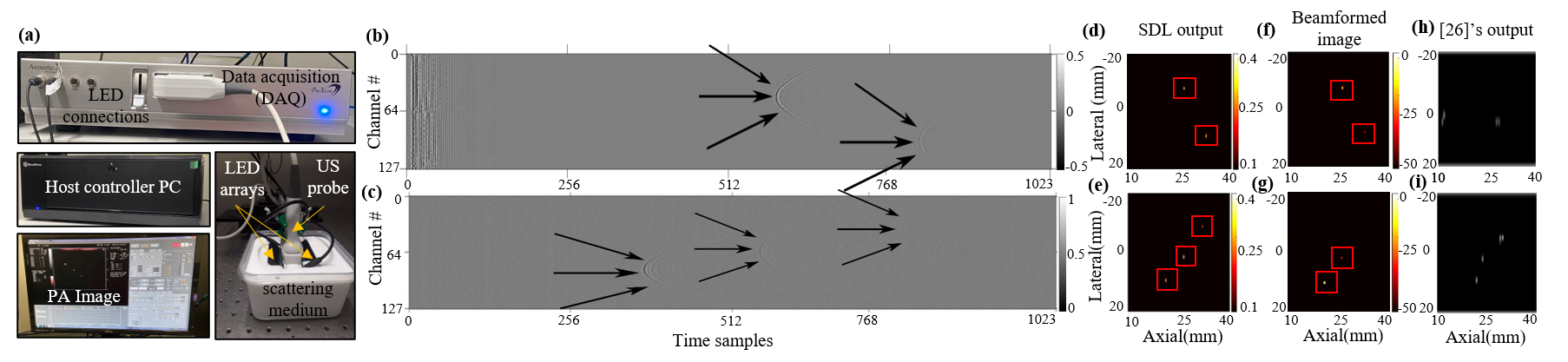}
    \caption{(\textbf{a}) The experimental setup to measure photoacoustic signals from light-absorbing pencil lead targets submerged inside an optically scattering intralipid solution. (\textbf{b}) Experimental photoacoustic data for two different experiments and their corresponding (\textbf{d},\textbf{e}) heatmap outputs from the network, (\textbf{f},\textbf{g}) respective beamformed images, \color{black} and (\textbf{h},\textbf{i}) \cite{b11}'s outputs (centers of bounding boxes are shown). }
    \label{fig:exp2}
\end{figure*}

\begin{figure}[!h]
\setlength\belowcaptionskip{-1\baselineskip}
    \centering
    \includegraphics[width=.5 \textwidth]{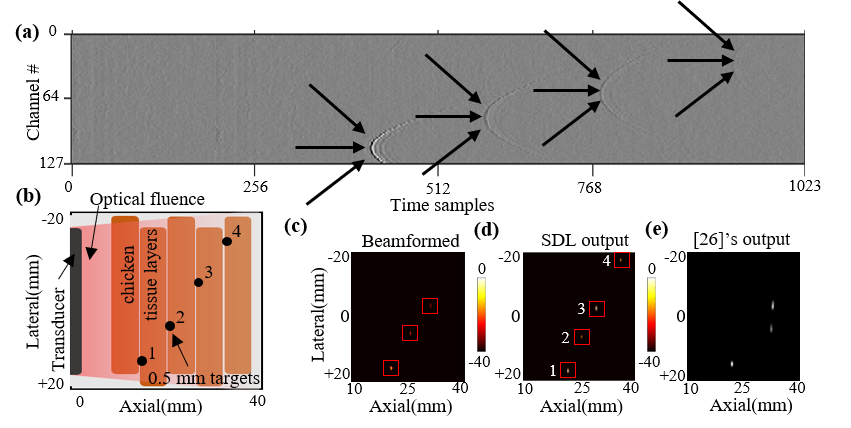}
    \caption{(\color{black}\textbf{a}) Photoacoustic sample, (\textbf{b}) the chicken tissue experimental setup used for capturing this sample, (\textbf{c}) the beamformed image, (\textbf{d}) SDL's output heatmap, and (\textbf{e}) [23]'s output (centers of the bounding boxes are shown).}
    \label{fig:chicken}
\end{figure}

\noindent \textbf{Ablation Study:}  To lend greater insight into SDL components,  Table \ref{tab:new_dataset} performs an ablation study by reporting results for two SDL variants: one without wavefront/noise filters (and hence also without the ${\mathcal L}_{WN}$ regularizer) and another single encoder-decoder configuration that does not employ the denoising decoder. The gains of SDL, over its counterparts that are stripped of regularization, are readily apparent in Table  \ref{tab:new_dataset} showing the benefit of each regularizer in Eq. (\ref{eq:finalLoss}). While the networks in state of the art were trained and designed for different experimental setups as investigated in Section III-C, the single encoder-decoder results in Table \ref{tab:new_dataset} essentially can be thought to represent the performance of \cite{b11,b13} because they are without the architectural enhancements and regularizers that SDL employs. Fig. \ref{fig:samples}i shows a test sample with a scattering level of 20 $cm^{-1}$ and Fig. \ref{fig:sim2} shows its corresponding cleaned output and heatmap predicted by the SDL network (trained over varying scattering levels), which effectively localizes multiple targets while beamforming in Fig. \ref{fig:samples}i does not detect all 4 targets.
 \begin{table*}[h!]
  \caption{The performance of three networks: SDL, the network without filters, and single decoder for different noise levels with models' training time (T1) and inference time (T2). Note that T2 is reported per image sample. The benefits of wavefront/noise filters when the training set is limited are clearly apparent with best results indicated in bold.}
        \centering
         \resizebox{\textwidth}{!}{
        \begin{tabular}{|c|c|c|c|c|c|c|c|c|c|}
        \hline
             \multirow{2}{*}{$\mu$'s}&\multirow{2}{*}{Architecture}&\multicolumn{3}{|c|}{Depth$<$35 mm}&\multicolumn{3}{|c|}{Depth$\geq$35 mm}&\multirow{2}{*}{\color{black}T1(hr)}&\multirow{2}{*}{\color{black}T2(sec)} \\
             &&Ax(\textmu m)&Lat(\textmu m)&Euc(\textmu m)&Ax(\textmu m)&Lat(\textmu m)&Euc(\textmu m)&&\\
            %  \hline
            %  1$cm^{-1}$&single decoder&4.60&4.56&7.86&5.03&5.12&8.50\\
             \hline
             \multirow{3}{*}{10 $cm^{-1}$}&single decoder&15.67&13.30&23.22&14.66&10.59&20.51&\color{black}8.29&\color{black}0.049\\
             &w/o wavefront/noise filters&3.48&4.35&6.74&3.18&2.28&4.37&\color{black}37.44&\color{black}0.051\\
             &SDL&\textbf{2.13}&\textbf{1.78}&\textbf{3.10}&\textbf{2.16}&\textbf{1.75}&\textbf{3.07}&\color{black}55.30&\color{black}0.053\\
             
             \hline
             \multirow{3}{*}{20 $cm^{-1}$}&single decoder&20.32&18.49&30.56&19.47&18.93&29.88&\color{black}8.25&\color{black}0.057\\
             &w/o wavefront/noise filters&5.99&4.22&7.91&4.05&2.53&5.27&\color{black}37.52&\color{black}0.046\\
             &SDL&\textbf{3.62}&\textbf{3.18}&\textbf{5.15}&\textbf{1.99}&\textbf{1.84}&\textbf{3.05}&\color{black}55.7&\color{black}0.050\\
             \hline
             \multirow{3}{*}{mixed}&single decoder&4.54&4,76&8.22&4.71&4.24&7.70&\color{black}15.65&\color{black}0.063\\
             &w/o wavefront/noise filters&1.24&0.68&1.57&1.20&0.80&1.59&\color{black}50.76&\color{black}0.068\\
             &SDL&\textbf{0.98}&\textbf{0.39}&\textbf{1.17}&\textbf{1.02}&\textbf{0.42}&\textbf{1.22}&\color{black}94.86&\color{black}0.065\\
             \hline
        \end{tabular}}
        \label{tab:new_dataset}
        \squeezeup
    \end{table*}\\
\color{black}\noindent \textbf{Cascaded vs. parallel denoising}: To emphasize  the benefits of our denoising approach over a cascaded strategy (denoising followed by localization), we compare SDL with a network consisting of a denoiser followed by a localization network. We make use of the denoising network proposed by Awasthi et al. \cite{b57} which is based on a hybrid U-Net \cite{b4} and represents state-of-the-art denoising for PA imaging applications. For the localization part, we incorporate the same encoder-decoder structure as the encoder and localization decoder in SDL. Both networks are trained over our combined training set (with all scattering noise levels) and tested on a set which is the combination of our test sets for scattering noise levels of 10 $cm^{-1}$ and 20 $cm^{-1}$ (Training size of 24,000 and test size of 4,000). Table \ref{tab:loc_cascaded} shows the localization results and Table \ref{tab:den_cascaded} shows the denoising performance of the networks in terms of peak signal to noise ratio (PSNR) and structural similarity index measure (SSIM) \cite{b58}. While the inference time is significantly less for SDL, it outperforms the cascaded network with respect to both localization and denoising. We can visually verify the quality of denoised outputs produced by SDL in Fig. \ref{fig:denoise}c compared to the output of the denoising network proposed in \cite{b57} - Fig. \ref{fig:denoise}d. We contend that SDL outperforms state of the art because, in our proposal, localization helps denoising (and vice-versa) via the shared encoder in Fig. \ref{fig:network}. 
\begin{table*}
 \color{black}
 \caption{\color{black}The localization results of the two models, SDL and the cascaded network for the combined 10 $cm^{-1}$ and 20 $cm^{-1}$ test set. T is the inference time reported per image sample.}
 \centering
         \resizebox{\textwidth}{!}{
        \begin{tabular}{|c|c|c|c|c|c|c|c|}
        \hline
             \multirow{2}{*}{Architecture}&\multicolumn{3}{|c|}{Depth$<$35 mm}&\multicolumn{3}{|c|}{Depth$\geq$35 mm}&\multirow{2}{*}{T(sec)} \\
             &Axial(\textmu m)&Lateral(\textmu m)&Euclidean(\textmu m)&Axial(\textmu m)&Lateral(\textmu m)&Euclidean(\textmu m)&\\
            %  \hline
            %  1$cm^{-1}$&single decoder&4.60&4.56&7.86&5.03&5.12&8.50\\
             \hline
             cascaded network&7.33&5.08&10.20&8.96&5.82&12.28&0.097\\
             SDL&\textbf{0.91}&\textbf{0.38}&\textbf{1.09}&\textbf{0.96}&\textbf{0.40}&\textbf{1.15}&\textbf{0.059}\\
             \hline
        \end{tabular}}
        \label{tab:loc_cascaded}
        \squeezeup
    \end{table*}
\begin{table}
\color{black}
\caption{\color{black} The denoising results of the two models, SDL and the cascaded network for the combined 10 $cm^{-1}$ and 20 $cm^{-1}$ test sets. SSIM is an image quality measure normalized to (0,1).}
\centering
 \resizebox{.35 \textwidth}{!}{
    \begin{tabular}{|c|c|c|}
    \hline
        Architecture&PSNR(dB)&SSIM  \\
        \hline
         cascaded network&39.00&0.9929\\
         \hline
         SDL&\textbf{42.36}&\textbf{0.9945}\\
         \hline
    \end{tabular}}
    \label{tab:den_cascaded}
    \squeezeup
\end{table}\\
\color{black}\noindent \textbf{Impact of the number of the elements:} To investigate the reliability of the network when the number of elements (transducers) decreases, we design a new experimental scenario with a dataset that is generated with the same setup as our practically representative simulated dataset except that the number of transducers decreases from 256 to 128. In this experiment, we generate 24,000 samples with optical scattering coefficients of 1, 10, and 20 $cm^{-1}$ ($\frac{1}{3}$ of the whole dataset each) with random number of targets (1-4). We used a set of size 18,000 for training and the 6,000 remaining samples for test. We train two models in this experiment: SDL and a variation of \cite{b13}'s network for detecting multiple targets (referred to as the single decoder model). We evaluate the performance of each model on the test set. Table \ref{tab:128} shows the results. As can be expected, when the number of elements decreases, a slight degradation is observed in the performance of both networks (SDL and single decoder). This is because the information that the network can grasp decreases with the reduced number of elements, {\em viz.} the model now has a downsampled lateral view of the wavefront. Note however in Table \ref{tab:128}, the relative benefits of SDL still remain.\\
The performances of the SDL network and \cite{b11} are evaluated with experimental PA data from two phantoms, intralipid optical scattering and a chicken breast tissue phantom, consisting of 0.5 mm pencil lead targets placed at different depths. Since the size of the experimental samples is different from the training data, the training data is resized and the samples with targets deeper than 35 mm are removed to mimic the experimental RF data and the network is trained over the modified dataset. The experimental samples and corresponding heatmaps are shown in Fig. \ref{fig:exp2}, \ref{fig:chicken}. Corresponding beamformed image for each sample shows the accuracy level of SDL in these highly challenging scenarios. Remarkably, SDL detects all three targets accurately in Fig. \ref{fig:exp2}e vs. two targets detected by the beamformed output in Fig. \ref{fig:exp2}g and the third target detected by \cite{b11} with lower accuracy. The SDL's superiority is more clear for the chicken tissue data in Fig. \ref{fig:chicken} where both beamformed image and \cite{b11} fail in detecting the 4th target.
\begin{table*}
 \color{black}
 \caption{\color{black}The performance results of the two models, SDL and single decoder for 128-element transducer dataset. The margin between the two model performances is almost the same as 256 elements dataset (mixed scattering noise level).}
 \centering
         \resizebox{\textwidth}{!}{
        \begin{tabular}{|c|c|c|c|c|c|c|c|}
        \hline
             \multirow{2}{*}{Architecture}&\multicolumn{3}{|c|}{Depth$<$35 mm}&\multicolumn{3}{|c|}{Depth$\geq$35 mm} \\
             &Axial(\textmu m)&Lateral(\textmu m)&Euclidean(\textmu m)&Axial(\textmu m)&Lateral(\textmu m)&Euclidean(\textmu m)\\
            %  \hline
            %  1$cm^{-1}$&single decoder&4.60&4.56&7.86&5.03&5.12&8.50\\
             \hline
             single decoder&6.73&5.18&9.40&6.94&5.58&9.91\\
             SDL&\textbf{1.96}&\textbf{2.53}&\textbf{3.49}&\textbf{2.05}&\textbf{2.51}&\textbf{3.51}\\
             \hline
        \end{tabular}}
        \label{tab:128}
    \end{table*}
\color{black}
\section{Discussions and Conclusion}
Design of custom denoising filters that affect the output coordinates by the autoencoder is achieved by sharing the low dimensional output of the encoder between the localization decoder and our novel denoising decoder. Experimental results indicate this method works by influencing the parameters of the encoder as the denoising decoder is not used in the test phase. Although the intuitive approach for training the localization decoder is to define the \textbf{mean square error} between predicted and ground truth target coordinates as the loss function (Eq. (\ref{eq:denoiser})), this approach is naive as it does not make use of any domain knowledge specific to the problem. We take advantage of our understanding of how noise and signal appear in the photoacoustic RF data by devising \textbf{Noise} and \textbf{Wavefront} filters and formulating custom regularizers to augment the loss function. Of note, how the network is trained is a matter of interest as well. Finding the optimal value for each of the regularization constants in the loss is not trivial and convergence of the training loss of the network with many regularization terms is not guaranteed. We experimented with different training routines and found that the best outcome was achieved by neglecting the wavefront and noise terms for the first few training epochs, allowing the network to converge to a sub-optimal point, and then engaging the noise and wavefront terms to drive the network toward a more optimal point.\\  
The experiments on the datasets of \cite{b11} and \cite{b13} prove the capability of our network in dealing with simulated additive noise in addition to signal attenuation due to optical scattering. This is due to the fact that our designed robustness to noise (using the Dec-D branch in Fig. \ref{fig:network}) is data adaptive and does not make any statistical assumptions on the nature of the noise. According to our results on the dataset of \cite{b11}  (Table \ref{tab:Comparison}), the percentage of test samples for which our network has a total error (mean Euclidean error) of less than 0.5 mm shows very slight degradation as the SNR decreases to -3dB, and this degradation becomes more noticeable as the noise level increases. This may be explained by considering the fact that the labels used for training the network are the bounding box centers provided by \cite{b11}, and the data used for training the network is the resized version of the original data. On the other hand, there is a surprising trend in the results of \cite{b11}, which is the unpredictable improvement when noise is added. Note that we exactly reproduced the results of \cite{b11} from their work, while training and testing on the same samples (as the authors of \cite{b11} kindly make their dataset available). \cite{b13}'s dataset, on the other hand, is more realistic with respect to the effect of optical scattering in photoacoustic imaging. However, it only considers single target data. Comparing the performance of the network in \cite{b13} with proposed SDL (Table \ref{Comparison_tuffc}), two observations can be made: 1) the overall performance is improved with respect to different criteria (lateral, axial, and Euclidean error) and 2) results for targets deeper than 40 mm indicate higher gains via SDL in both lateral and axial error. This deep region is where optical scattering plays a dominant role, as the weakened optical fluence results in photoacoustic signals that barely peak above the noise floor in the data. For real-world experimentally captured data with an optical scattering coefficient of 20 $cm^{-1}$, Fig. \ref{fig:exp1} confirms a highly valuable benefit of SDL over state of the art in handling significant levels of scattering noise practically. We attribute this success to the explicit attention to noise robustness in the design of SDL via custom-designed regularizers.
Overall, the proposed architecture shows flexibility with respect to noise level/type and number of targets, which suggests its potential significance for PA imaging applications including cancer detection and treatment \cite{b22,b23,b24}, and treatment of vascular diseases such as deep vein thrombosis \cite{b25} and blood vessel morphology \cite{b26}. One practical constraint in the application of deep learning methods for many medical imaging problems is that representative labeled training data is often not abundant, unlike analogous detection and classification problems in consumer imaging; PAI shares this challenge. The incorporation of problem specific domain knowledge via regularization terms/architectural innovations, as have been performed in this work, can be useful in addressing the challenge of limited training. Our investigation in Fig. 7 demonstrates that SDL can outperform state of the art while using half the training data.\\
In conclusion, this work reveals the importance of taking into account the optical scattering noise for photoacoustic target localization problems. Deep learning frameworks have been used before for this problem with promising success, but those architectures did not explicitly build or enhance noise robustness. Existing deep learning approaches for PAI also rely significantly on the quantity and quality of training data available.
Our proposal addresses these challenges by exploiting the characteristics of PA images towards a noise-robust approach. Specifically, a shared encoder-dual decoder architecture is designed for simultaneous denoising and localization. Custom-designed
regularizers inspired by the shape of the noise and the signal in the PA images help fit the reconstructed data to the ground truth noiseless data more effectively. These regularizers also help enhance performance when training data samples are limited. Finally, we design a new dataset that introduces significant diversity with respect to the scattering noise levels and the number of the targets. Experiments performed on the practically representative simulated dataset, as well as existing simulated and experimental datasets from the state of the art, demonstrates the capability of our proposed deep network in detecting targets with higher accuracy. Our method is also shown to successfully operate in scenarios where existing trained networks do not produce meaningful outputs. \color{black} While our proposed methodology benefits from custom priors to effectively improve the signal strength, one can design the wavefront and noise filters using a generative model such as GANs to potentially achieve even better prior inspired regularization in Eq (\ref{eq:finalLoss}). GAN-inspired learned filters may hence enhance both denoising and localization and form a viable future research direction. \color{black} Moreover, our model is designed for the detection of up to 4 targets in photoacoustic data, which could be a limit for the cases when the number of targets is more. However, we can modify the output layer of the network to detect a higher number of targets. A versatile detection of an arbitrary number of targets is a future research direction. Also, a broad direction for all machine learning efforts in PA imaging is the detection of non-point targets.
\color{black}
\section{Acknowledgement}
The authors would like to thank \textit{Isaac Gerg} for proofreading the paper and helping increase the comprehensibility of the manuscript. 
\appendices
%section*{Acknowledgment}
%\newpage
%\section*{References and Footnotes}
\typeout{} 
\bibliographystyle{IEEEtran}
\bibliography{tmi.bib}

\end{document}